\documentstyle[preprint,aps,eqsecnum]{revtex}
\tightenlines
\def\m@thcombine#1#2{%
  \setbox0=\hbox{$#1$}
  \setbox1=\hbox{$#2$} 
  \ifdim\wd0>\wd1
    \setbox0=\hbox to\wd1{\hss\box0\hss}
  \else
    \setbox1=\hbox to\wd0{\hss\box1\hss}
  \fi
  \mathop{\vcenter{
    \offinterlineskip\box0\box1}}}
\def\lesim{\m@thcombine<\sim}
\def\gesim{\m@thcombine>\sim}

\begin{document}

\draft
\title{ VACUUM INSTABILITY IN THE ABELIAN HIGGS MODEL WITH STRINGS }

\author{V. Gogohia, M. Priszny\'ak}

\address{HAS, CRIP, RMKI, Theory Division, P.O.B. 49, H-1525 Budapest 114, Hungary \\ email addresses: gogohia@rmki.kfki.hu and prisz@sunserv.kfki.hu}                                   

\maketitle

\begin{abstract}
Using the effective potential approach for composite operators, we have analytically evaluated the truly nonperturbative vacuum energy density in the Abelian 
Higgs model of dual QCD ground state.  This quantity is defined as
integrated out of the truly nonperturbative part of the full gluon propagator  
over the deep infrared region (soft momentum region). Defined in this way it is
manifestly gauge invariant. We have explicitly shown that the corresponding   
effective potential always has an imaginary part. This means
that the vacuum of this model with string contributions is unstable
against quantum corrections.  
\end{abstract}


\vfill

\eject

\section{Introduction}

In our previous publications [1,2], we have formulated a general method how to 
correctly calculate the truly nonperturbative vacuum energy density (VED) in the QCD quantum models of its ground state using the effective potential approach
for composite operators [3-5]. The truly nonperturbative VED is defined as
integrated out of the truly nonperturbative part of the full gluon propagator over the deep infrared (IR) region (soft momentum region). The nontrivial minimization procedure which can be done only by the two different ways (leading however to the same numerical value (if any) of the truly nonperturbative VED) 
makes it possible to determine the value of the soft cutoff in terms of the corresponding nonperturbative scale parameter which is inevitably present in any
nonperturbative model for the full gluon propagator. If the chosen Ansatz for the full gluon propagator is a realistic one, then our method uniquely determines the truly nonperturbative VED, which is always finite, automatically negative
and it has no imaginary part (see, for example Refs. [2,6,7]).
Our method can serve as a test of different QCD quantum as well as classical vacuum models since it provides an exact criterion for the separation "stable vs.
unstable" vacuum. The vacuum stability in the classical models is also important. In the above-mentioned paper [1], we have already shown that the vacuum of the Abelian Higgs model [8,9] without string contributions is unstable against quantum corrections. The main purpose of this work is to investigate the vacuum structure of the Abelian Higgs model on account of string contributions.    
      
 The relevant expression for the truly nonperturbative Yang-Mills (YM) VED, which was obtained in Ref. [1], is (four-dimensional Euclidean apace)             
                                                     
\begin{equation}
\epsilon_g^{np} = - { 1 \over \pi^2} \int_0^{q_0^2} dq^2 \ q^2 
\left[ \ln \left(1 + q^2 F^{NP}(q^2) + G^{NP}(q^2) \right) - \left( q^2 F^{NP}(q^2) + G^{NP}(q^2) \right) \right], 
\end{equation}
where $q^2_0$ is the above-mentioned soft cutoff. The truly nonperturbative gluon form factors $F^{NP}(q^2)$ and $G^{NP}(q^2)$ are defined as follows: 

\begin{eqnarray}
F^{NP}(q^2, \Lambda_{NP}) &=& F(q^2, \Lambda_{NP})- F(q^2, \Lambda_{NP} = 0),  
\nonumber\\
G^{NP}(q^2, \Lambda_{NP}) &=& G(q^2, \Lambda_{NP})- G(q^2, \Lambda_{NP} = 0),
\end{eqnarray}
which explains the difference between the truly nonperturbative and the full gluon form factors which are nonperturbative themselves, while $\Lambda_{NP}$ is precisely the scale parameter responsible for nonperturbative dynamics in
the model under consideration. Moreover, it guaratees that the truly nonperturbative VED (1.1) is manifestly gauge  
invariant quantity. Though the full gluon propagator is explicitly gauge dependent, its truly nonperturbative part is not since the explicit gauge          
dependence on the parameter which fixes the direction of gauge  goes away along
with the perturbative terms contained in $F,G(q^2, \Lambda_{NP} =0) \equiv F^{PT},G^{PT}(q^2)$. It is easy to understand that by "PT" we mean intermediate (IM) plus    
ultraviolet (UV) regions, i.e., "PT= IM + UV" (the IM region remains $terra \  
incognita$ in QCD). Fortunately the "PT" part is of no importance here.        
                                                      
Thus the separation of "NP vs. PT" becomes exact because of the definition     
(1.2). The separation of "soft vs. hard" momenta also becomes exact because of 
the above-mentioned minimization procedure. The analysis of the truly nonperturbative VED (1.1) after the scale factorization provides in addition   
an exact criterion for the separation of "stable vs. unstable" vacuum.     
Thus the truly nonperturbative VED as it is given in Eq. (1.1) is uniquely
defined. It is truly nonperturbative since it contains no even one bit of perturbative information, i.e., it is free from all kinds of perturbative contamination. This is absolutely similar to lattice approach where by using different 
"smoothing" techniques such as "cooling" [10], "cycling" [11], etc. it is possible to "wash out" all types of the perturbative fluctuations and excitations of
the gluon field configurations from the QCD vacuum in order to deal only with  
its true nonperturbative structure.                                            

The above-briefly-described general method [1,2] can serve as a test of QCD vacuum different not only quantum, classical but lattice models as well.
Let us formulate now the classical Abelian Higgs model of 
the dual QCD ground state [1,8,9].

\section{Abelian Higgs model}

In the dual Abelian Higgs theory which confines electric charges the coefficient functions $F(q^2)$ and $G(q^2)$, which determine the vacuum of this model,  
 are [1,8] (Euclidean metrics)  

\begin{eqnarray}
F(q^2) &=& { 1 \over q^2 + M^2_B } \left( 1 + { M^4_B D^{\Sigma}(q^2) \over q^2 + M^2_B} \right) , \nonumber\\                                                
G(q^2) &=& - { M^2_B \over q^2 + M^2_B } \left( 1 - M^2_B { q^2 D^{\Sigma}(q^2) \over q^2 + M^2_B} \right),
\end{eqnarray}
where $M_B$ is the mass of the dual gauge boson $B_{\mu}$ and $D^{\Sigma}(q^2)$
represents the string contribution into the gauge boson propagator. The mass scale paprameter $M_B$ is the scale responsible for nonperturbative dynamics in this model (in our notations $\Lambda_{NP} = M_B$). When it formally goes to zero, then one recovers the free perturbative expressions indeed,  $q^2F^{PT}(q^2) =1$ and $G^{PT}(q^2) =0$. Removing 
the string contributions from these relations we get

\begin{equation}
F^{no-str.}(q^2) = { 1 \over q^2 + M^2_B }, \qquad                      
G^{no-str.}(q^2) = - { M^2_B \over q^2 + M^2_B },
\end{equation}
i.e., even in this case these quantities remain nonperturbative. 
The truly nonperturbative counterparts of the coefficient functions (2.1) because of the definitions (1.2) are

\begin{eqnarray}
F^{NP}(q^2) &=& - { M_B^2 \over q^2 (q^2 + M^2_B) } \left( 1 - { M^2_B q^2 D^{\Sigma}(q^2) \over q^2 + M^2_B} \right) , \nonumber\\                           
G^{NP}(q^2) &=& - { M^2_B \over q^2 + M^2_B } \left( 1 - { M^2_B q^2 D^{\Sigma}(q^2) \over q^2 + M^2_B} \right),
\end{eqnarray}
while with no-string contributions they become

\begin{equation}
F^{no-str.}_{NP}(q^2) = - { M^2_B \over q^2 (q^2 + M^2_B) }, \qquad     
G^{no-str.}_{NP}(q^2) = - { M^2_B \over q^2 + M^2_B }.
\end{equation}
Both expressions (2.3) and (2.4) are truly nonperturbative indeed, since they become zero in the perturbative limit ($M_B \longrightarrow 0$), when only perturbative phase remains. From these relations also follows

\begin{eqnarray}                           
G^{NP}(q^2) &=& q^2 F^{NP}(q^2) = - { M^2_B \over q^2 + M^2_B } \left( 1 - {M^2_B q^2 D^{\Sigma}(q^2) \over q^2 + M^2_B} \right), \nonumber\\
G^{no-str.}_{NP}(q^2) &=& q^2 F^{no-str.}_{NP}(q^2) = - { M^2_B \over q^2 + M^2_B },
\end{eqnarray}
so the truly nonperturbative vacuum energy density (1.1) will depend on  
the one function only, say, $G^{NP}(q^2)$ (see next section).

Although the expressions (2.1) for the gluon propagator are
exact, nevertheless they contain an unknown function $D^{\Sigma}(q^2)$ which is
the intermadiate string state contribution into the gauge boson propagator [8].
It can be considered as a glueball state with the photon quantum numbers $1^-$.
The bahavior of this function $D^{\Sigma}(q^2)$ in the IR region ($q^2 \rightarrow 0$) can be estimated as follows [8]: 

\begin{equation}                           
D^{\Sigma}(q^2) = {C \over q^2 + M^2_{gl}} + ..., 
\end{equation}
where $C$ is a dimensionless parameter and $M^2_{gl}$ is the mass of the lowest
$1^-$ glueball state. The dots denote the contributions of heavier states. 
Thus, according to Eqs. (2.1) and (2.6), the coefficient functions in the IR limit behave like

\begin{equation}                           
F(q^2) = {1 \over M^2_B} + {C \over M^2_{gl}} + O(q^2), \qquad
G(q^2) = - 1 +  O(q^2), \qquad  q^2 \rightarrow 0.    
\end{equation}
At the same time according to Eqs. (2.3), (2.5) and (2.6) their truly nonperturbative counterparts behave like

\begin{equation}
G^{NP}(q^2) = q^2 F^{NP}(q^2) =  - 1 + O(q^2), \ q^2 \rightarrow 0,     
\end{equation}
i.e., in the same way as $G(q^2)$ in Eq. (2.7). Comparing with Eq. (2.4), one can can conclude in that the string contribution is not dominant in the most important nonperturbative infrared (IR) region. It is the next-to-leading order term in the IR. In the next section, we will show that precisely this feature leaves the vacuum of this model unstable even on account of string contributions  
into the gauge boson propagator.

\section{Vacuum structure in the Abelian Higgs model}

Let us calculate the truly nonperturbative VED in the Abelian
Higgs model described in the preceding section. Because of the relations (2.5),
it depends on the one structure function only, namely                          
       
\begin{equation}
\epsilon_g^{np} = { 1 \over \pi^2} \int_0^{q_0^2} dq^2 \ q^2 
\left[ 2 G^{NP}(q^2) - \ln \left(1 + 2 G^{NP}(q^2) \right) \right]. 
\end{equation}
As was emphasized in our paper [1], it is always convenient to factorize the dependence on a scale in the truly nonperturbative VED (3.1). The full gluon form
factors always contain at least one scale parameter responsible for the nonperturbative dynamics in the model under consideration, $\Lambda_{NP}$. Within our general method 
we are considering it as free one, i.e., as "running" (when it formally goes 
to zero, only perturbative phase survives in the model) and
its numerical value (if any) will be used only at final stage in order to numerically evaluate the corresponding truly nonperturbative VED (if any).
We can introduce dimensionless variables and parameters by
using completely extra scale (which is aways fixed in comparison with $\Lambda_{NP}$), for example flavorless QCD asymptotic scale parameter $\Lambda_{YM}$ (the so-called A-scheme of Ref. [1]). However, in this case it is much more convenient to use the B-scheme [1] with $\Lambda_{NP} = M_B$ in order to factorize the scale dependence in the expression (3.1)
, namely

\begin{equation}
z={q^2 \over M^2_B}, \quad  z_0={q_0^2 \over M^2_B},  \quad  a={M^2_{gl} \over M^2_B}.  
\end{equation}
Introducing further the effective potential at a fixed soft cutoff [1], one obtains  

\begin{equation}
\bar \Omega_g (z_0) = { 1 \over q_0^4} \epsilon_g^{np
}(z_0)= {1 \over \pi^2}  z_0^{-2}                                              
\int_0^{z_0} dz \ z \left[ 2 G^{NP}(z)  -  \ln \left( 1 + 2 G^{NP}(z) \right) \right],
\end{equation}
where the gluon form factor $G^{NP}(z)$ (2.5) obviously becomes  

\begin{equation}
G^{NP}(z)= - {1 \over 1+z} \left[ 1 - {C z \over (1+z) (z+a)} \right].
\end{equation}
Integrating Eq. (3.3) on account of (3.4), one obtains 

\begin{equation}
\bar \Omega_g (z_0) = - { z_0^{-2} \over 2 \pi^2} \left[ {2 z_0 (a-1-C) \over(a-1)} - {2Ca^2 \over (a-1)^2} \ln \left(1 + {z_0 \over a}\right)   
- 2 \left( 1 + {C(1-2a) \over (a-1)^2 } \right) \ln (1 + z_0)  + I (z_0) \right],
\end{equation}
where

\begin{equation}
I(z_0) =                                               
\int_0^{z_0} dz \ z \ln \left[ 1 - {2 \over 1+z} \left( 1 - {C z \over (1+z) (z+a)} \right) \right].
\end{equation}
When $C=0$ one recovers the effective potential without string contributions obtained in Ref. [1]. 

The specific feature of this model is that the combination 

\begin{equation}
\nu = {z_0 \over a} ={q_0^2 \over M^2_{gl}}  
\end{equation}
is fixed when the soft cutoff $q_0$ and $M_{gl}$ are fixed. Precisely this takes place in our case ($q_0$ is fixed bacause of (3.3) and  $M_{gl}$ is fixed $a \ priori$). Thus the effective potential (3.5) becomes 

\begin{equation}
\bar \Omega_g = { z_0^{-2} \over 2 \pi^2} \left[ {2Cz_0^2 \ln (1 + \nu) \over (z_0-\nu)^2} - {2 z_0 (z_0-\nu-C\nu) \over(z_0-\nu)}    
+ 2 \left( 1 + {C\nu(\nu-2z_0) \over (z_0-\nu)^2 } \right) \ln (1 + z_0) -   
I (z_0) \right],
\end{equation}
where now $\bar \Omega_g \equiv \bar \Omega_g (z_0, \nu, C)$, i.e., in fact the
dependece on $z_0$ is more complicated. It becomes a function of the three independent above-indicated variables and       

\begin{equation}
I(z_0) \equiv I(z_0, C, \nu) =                                               
\int_0^{z_0} dz \ z \ln \left[ {-1+z \over 1 + z} + {2 C z \nu \over (1+z)^2 (z_0 + \nu z)} \right].
\end{equation}

The bahaviour of the effective potential (3.8) with respect to the parameters  
$C$ and $\nu$ is not restricted at all, while with respect to the soft cutoff
$z_0$ it should vanish at infinity ($z_0 \rightarrow \infty$) since the truly nonperturbative VED vanishes
when the perturbative phase remains only in the perturbative ($M_B \rightarrow 
0$) limit. Indeed, asymptotics of the effective potential (3.8) to-leading order are

\begin{equation}
\bar \Omega_g (z_0)_{z_0 \rightarrow 0} \sim {1 \over \pi^2} \left[
{ \ln (1 + \nu - 2C \nu) \over 2 \nu^2 (2C -1)^2}  -   
{ 9 C \over \nu} - { 1 -4C- \nu^2 \over 2 \nu^2} \ln (1 + \nu) -               
{1 \over 2} \ln (2C \nu - 1 - \nu) \right]
\end{equation}
and

\begin{equation}
\bar \Omega_g (z_0)_{z_0 \rightarrow \infty} \sim {2 \over \pi^2} z_0^{-2} \ln z_0.
\end{equation}
Thus in the perturbative limit ($z_0 \rightarrow \infty$) it vanishes as it should be. At the same time, from asymptotic behavior (3.10) it follows that at   
any values of the parameters $C$ and 
$\nu$, the effective potential (3.8) at zero point ($z_0=0$) will always contain the imaginary part which manifests a possible vacuum instability [12] in this
model.    

  However, it can be proven that the effective potential (3.8) will always have
an imaginary part at any finite values of its variables, $C, \ \nu$ and 
$z_0$ as well.
To this end, it suffices to investigate the integral (3.9), more precisely 
the function under logarithm, namely   

\begin{equation}
R \equiv R(z, z_0, C, \nu) =                                               
 {-1+z \over 1 + z} + {2 C z \nu \over (1+z)^2 (z_0 + \nu z)}.
\end{equation}
 Let us notice that $R(z=0, z_0, C, \nu)= -1$ holds true for $any \ fixed$ $C$,
$\nu$ and $z_0$. Since $R$ is regular as a function of $z$ in the whole interval $[0, z_0]$ for any $z_0$, $C$ and $\nu$\footnote{The poles at $z=-1$ and     
$z=-(z_0 / \nu)$ do not lie in the interval $[0, z_0]$ since $z_0$ is always positive, by definition.}, it simply follows from the Boltzano-Weierstrass       
theorem that there exists 
an interval (with $z=0$ as the left end point) where $R$ is negative, provided 
$R$ becomes non-negative somewhere in the interval $[0, z_0]$. If this were not
true then $R$ must be negative in the whole interval  $[0, z_0]$. Having such  
an interval where $R<0$, and taking
into consideration that the logarithm is a monotonous function, we certainly have an imaginary part in the effective potential (3.8) at any finite set of parameters, $z_0, \ C, \ \nu$.  

Thus one conludes in that the vacuum of the Abelian Higgs model is unstable against quantum corrections. Moreover, the string contributions
cannot cure this fundamental defect since (as mentioned above) they do not
rearrange the structure of the vacuum in the deep IR (nonperturbative) domain. 
  
One of the authors (V.G.) is grateful to M. Plikarpov for correspondence. We would like also to thank Gy. Kluge for discussion and useful remarks. This work was supported by the OTKA grant No: T016743.

\vfill

\eject

\end{document}